\documentclass[11pt]{amsart}
\usepackage{geometry}                
\usepackage{graphicx}
\usepackage{amssymb}
\usepackage{epstopdf}
\usepackage{hyperref}
\pdfoutput=1
\title{Deceleration Driven Wetting Transition Druing "Gentle" Drop Depostion}
\author{Hyuk-Min Kwon, Adam T. Paxson, Kripa K. Varanasi \\
Massachusetts Institute of Technology, Cambridge, MA\\
Neelesh A. Patankar\\
Northwestern University, Evanston, IL}

\begin{document}
\pagestyle{empty}
\maketitle
\begin{abstract}
We present high speed video of Cassie-Baxter to Wenzel drop transition during gentle deposition of droplets where the modest amount of energy is channeled via rapid deceleration into a high water hammer pressure. 
\end{abstract}
\section{Brief Explanation of Video Submission}
Roughness-induced superhydrophoic surfaces have been widely studied, and is well known that there are two wetted states on those surfaces. In one of the states, droplets reside on top of roughness features, i.e. in a Cassie-Baxter (CB) state$^1$. Droplets that impale the roughness grooves, i.e., in a Wenzel state$^2$, represent another commonly observed scenario. Recent experimental work has successfully revealed pressure-induced transition from the CB to the Wenzel state on rough hydrophobic substrates with pillar geometries. For a droplet to remain in the CB state, the transition-inducing wetting pressures must be less than the anti-wetting pressure which is the capillary pressure in the case of a textured surface. In the case of a Laplace pressure-induced transition, a smaller droplet will more readily transition to a Wenzel state. Another mechanism of transition, driven by gravity, was implicated by Yoshimitsu et al.$^3$ They found that larger droplets, above a critical size, transitioned to the Wenzel state. This result is opposite to the Laplace pressure-induced transition. This is surprising because the water droplets used in their experiments were 1$-$12 mg, where gravity is not expected to play a dominant role if the droplets are deposited gently. Usually, gravity is expected to be comparable to or larger than the surface tension forces for water droplets 82 mg or larger. It has remained unclear if these data are repeatable, or, if repeatable, the details of the transition process are unclear. 

It is found that if deposited quasi-statically, the CB droplet does not undergo gravity-driven transition to a Wenzel state, contrary to the conclusions of Yoshimitsu et al.$^3$ However, the high-speed video we report here reveals that the wetting transition is sensitive to perturbations that are unintentionally introduced during the deposition process. The consequent rapid deceleration of the droplet while it settles to its equilibrium state can induce transition to a Wenzel state. This transition occurs more readily for larger droplets. 

\section{Video Submissions}
\href{/APSpackages/anc/WettingTransition_GFM_2010.mp4}{Video 1} is a presentation-quality version of the fluid dynamics video. 

\href{/APSpackages/anc/WettingTransition_GFM_2010_for_Web.mp4}{Video 2} is a web-quality version of the fluid dynamics video. 

\section{References}
1.	Cassie, A.B.D. and S. Baxter, Wettability of porous surfaces. Transactions of the Faraday Society, 1944. 40: p. 546-551.

2.	Wenzel, R.N., Surface roughness and contact angle. Journal of Physical and Colloid Chemistry, 1949. 53(9): p. 1466-1467.

3.	Yoshimitsu, Z., A. Nakajima, T. Watanabe, and K. Hashimoto, Effects of surface structure on the hydrophobicity and sliding behavior of water droplets. Langmuir, 2002. 18(15): p. 5818-5822.
\end{document}